\documentclass[iop]{emulateapj}
\usepackage{apjfonts}

\catcode`\@=11
\catcode`\@=12
\shorttitle{EXPANSION OF YOUNGEST GALACTIC SNR G1.9+0.3}

\begin{document}

\title{Nonuniform Expansion of the  
Youngest Galactic Supernova Remnant G1.9+0.3}

\author{Kazimierz J. Borkowski,\altaffilmark{1}
Stephen P. Reynolds,\altaffilmark{1}
David A. Green,\altaffilmark{2}
Una Hwang,\altaffilmark{3}
Robert Petre, \altaffilmark{4}
Kalyani Krishnamurthy, \altaffilmark{5}
\&\ Rebecca Willett \altaffilmark{6}
}

\altaffiltext{1}{Department of Physics, North Carolina State University, 
Raleigh, NC 27695-8202; kborkow@unity.ncsu.edu} 
\altaffiltext{2} {Cavendish Laboratory; 19 J.J. Thomson Ave., 
Cambridge CB3 0HE, UK}
\altaffiltext{3}{Department of Astronomy, University of Maryland, 
College Park, MD 20742}
\altaffiltext{4}{NASA/GSFC, Code 660, Greenbelt, MD 20771}
\altaffiltext{5}{Department of Electrical and Computer Engineering, 
Duke University, Durham, NC 27708}
\altaffiltext{6}{Department of Electrical and Computing Engineering, 
University of Wisconsin-Madison, Madison, WI 53706}

\submitted{Accepted for publication in ApJ Letters}

\begin{abstract}

We report measurements of X-ray expansion of the youngest Galactic supernova
remnant, \object{G1.9+0.3}, using {\sl Chandra}
observations in 2007, 2009, and 2011.
The measured rates strongly 
deviate from uniform expansion, decreasing radially by about 
60\%\ along the X-ray bright 
SE-NW axis from $0.84\% \pm 0.06$\%\ yr$^{-1}$ to 
$0.52\% \pm 0.03$\% yr$^{-1}$. This corresponds to undecelerated ages of 
120 -- 190 yr, confirming the young age of \object{G1.9+0.3}, and implying 
a significant 
deceleration of 
the blast wave. The synchrotron-dominated X-ray emission  
brightens at a rate of $1.9\% \pm 0.4$\%\ yr$^{-1}$.
We identify bright outer and inner rims with the blast wave and reverse
shock, respectively.
Sharp density gradients in either ejecta or ambient
medium are required to produce the sudden deceleration of the reverse
shock or the blast wave implied by the large spread in expansion ages.
The blast wave could have been decelerated recently by an
encounter with a modest density discontinuity in
the ambient medium, such as found at a wind termination shock,
requiring strong mass loss in the progenitor.
Alternatively, the reverse shock
might have encountered
an order-of-magnitude  
density
discontinuity within the ejecta, such as found in pulsating
delayed-detonation Type Ia models. 
We demonstrate that the blast wave is much
more decelerated than the reverse shock in these models for 
remnants at ages similar to \object{G1.9+0.3}. Similar effects may also 
be produced by dense shells possibly 
associated with high-velocity features in Type Ia spectra.
Accounting for the asymmetry of \object{G1.9+0.3} will require
more realistic 3D Type Ia models.
\end{abstract}


\section{Introduction}
\label{intro}

\object{G1.9+0.3} is the remnant of the Galaxy's most recent supernova
\citep[][]{reynolds08}. Expansion between 2007 and 2009 of
about 1.6\% measured with {\sl Chandra} \citep[][]{carlton11}
gives an expansion (undecelerated) age of about 160 yr, but the
estimated mean expansion index $m$ ($R \propto t^m$) of about $m =
0.7$ gives an age $\sim 100$ yr.  The X-ray spectrum is dominantly
synchrotron with high absorption \citep{reynolds09}; 
lines of Fe, Si, and S are found in
small regions, with spectroscopic velocities of about 14,000 km
s$^{-1}$ \citep[][]{borkowski10}, consistent with the proper
motions for a distance of order 10 kpc \citep{roy14}.  For an assumed
location near the Galactic Center ($d = 8.5$ kpc), the diameter is
about 2 pc.  \object{G1.9+0.3} is quite asymmetric (Figure~1), and the
thermal emission is also asymmetrically distributed 
\citep[][]{borkowski13b}.  The relative prominence of Fe lines, high
velocities, absence of a pulsar-wind nebula, and bilaterally symmetric
synchrotron emission (as in SN 1006) all point toward a Type Ia
origin.  Only a very unusual core-collapse event could reproduce the
observations, while a reasonable SN Ia model can reach the observed
size and velocity for a mean external density of about 0.02 cm$^{-3}$
(Paper V).  Observing the rapid evolution of \object{G1.9+0.3} in
morphology and brightness can provide unprecedented information on the
dynamics of SN ejecta and on particle acceleration.

\begin{figure}
\epsscale{1.0}
\vspace{0.1truein}
\plotone{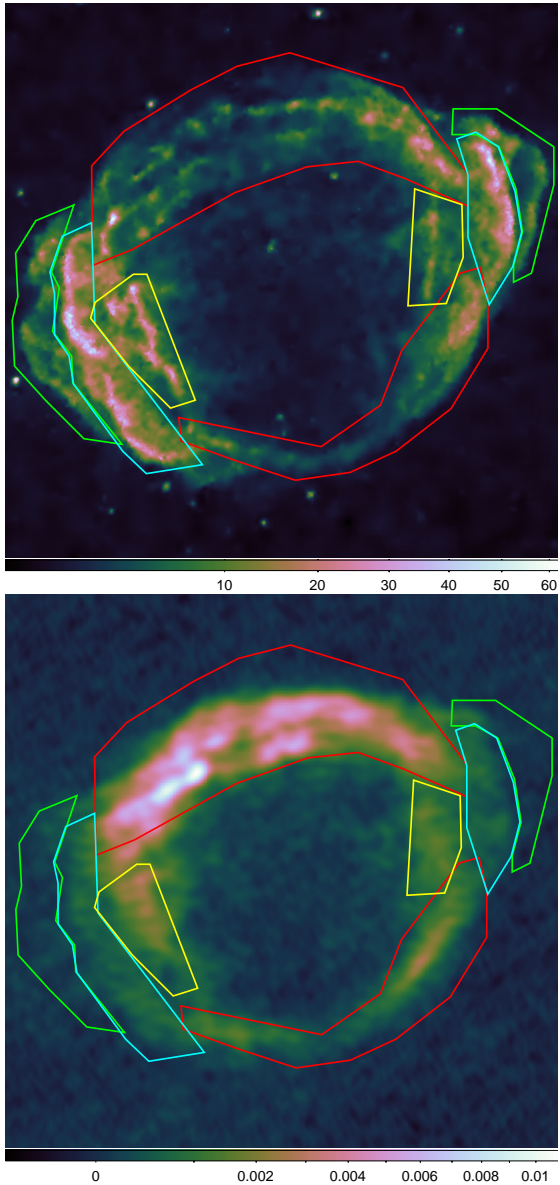}
\caption{Top: 
2011 {\sl Chandra} image of G1.9+0.3. Scale is in 
counts per ACIS pixel in the 1.2--8 keV energy range \citep[image was smoothed 
with the multiscale partitioning method of][]{krishnamurthy10}. 
Bottom: 1.4 GHz VLA radio image from 2008 December. 
Scale is in Jy beam$^{-1}$. Resolution $2.3'' \times 1.4''$. 
N is up and E is to the left.  Intensities shown with the cubehelix color 
scheme of \citet{green11}. Pairs 
of regions chosen for expansion studies are overlaid: outer (green), middle 
(magenta), and inner (yellow) pairs along the SE-NW axis, and NS (red) pair.  
Image size $123'' \times 123''$. 
\label{2xim}}
\end{figure}

\section{Observations}
\label{obssec}

{\it Chandra} observed \object{G1.9+0.3} with the ACIS S3 chip on
three epochs: (I) 2007 February and March, (II) 2009 July, and (III)
2011 May and July (details in Papers I and III -- VI). 
All observations were
done in Very Faint Mode, and reprocessed with CIAO v4.6 and CALDB
4.5.9. 
The corresponding effective exposure times are 49.6
ks, 237 ks, and 977 ks. Exposure-weighted time intervals between the
deep Epoch III and shorter Epoch I and II observations are 4.274 yr
and 1.861 yr. Alignment of observations at different epochs is
performed simultaneously with expansion measurements. 

Images, $512^2$ pixels in size, were extracted from the merged event files by
binning event positions to half the ACIS pixel size, so one image
pixel is $0.246'' \times 0.246''$. We also extracted data cubes,
$512^2 \times 64$ in size, using the same spatial pixel size, while
spectral channels from 84 to 595 (1.2--8.7 keV energy range) were
binned by a factor of 8.

X-ray spectra were extracted from individual rather than merged event
files, and then summed (response files were averaged).
Spectral analysis was performed with XSPEC v12.8.1 \citep{arnaud96},
using C-statistics \citep{cash79}. 
Background was modeled rather than subtracted. 
Spectra of \object{G1.9+0.3} were modeled with an
absorbed power law, using solar abundances of \citet{grsa98} in the
{\tt phabs} absorption model.  Thermal emission contributes
negligibly to broadband fluxes.

\section{Expansion} \label{expansion}

We first measured the overall expansion using the method described in
Paper V. Briefly, we smoothed 
the 2011 {\sl Chandra} data
cube with the spectro-spatial method of \citet{krishnamurthy10},
and summed the smoothed cubes in the spectral dimension to arrive at
smoothed images in the 1.2--8 keV energy range. (Softer X-rays 
are absorbed by the intervening interstellar 
medium, while the background dominates at the highest 
energies.) 
We used the smoothed image as a model for the surface brightness of
\object{G1.9+0.3} at Epoch III (Figure \ref{2xim}). This model
image was background-subtracted then fit to the unsmoothed
1.2--8 keV images from earlier epochs (i.e., shrunk to fit) using 
C-statistics.  Seven point sources within \object{G1.9+0.3} were 
masked out.  
There are four free parameters in this
model: a physical scaling factor, a surface-brightness scaling factor,
and expansion center coordinates.
Independent fits to 
Epoch I and
II images were consistent with the constant expansion. 
We then assumed the same expansion rate
while fitting jointly to the Epoch I and II images, but allowed for
independent surface-brightness scaling factors and expansion center
coordinates. Results of all measurements are listed in 
Table \ref{expansionfluxrates}. The
measured expansion rate is $0.589\% \pm 0.016$\%\ yr$^{-1}$ (all errors
are 90\%\ confidence), in reasonable agreement
with our previous measurement of $0.642\% \pm 0.049$\%\ (Paper
V). Within errors, the mean surface brightness does not vary with
time, again in agreement with Paper V.

To search for deviations from uniform expansion, we
systematically examined motions of large, distinct spatial structures.
The final regions chosen are overlaid over
X-ray and radio images in Figure \ref{2xim}. They include a pair on
the N and S rim and three pairs of regions along the major SE-NW axis:
an outer pair (``ears''), a middle pair containing the brightest rim
emission, and an inner pair with distinct interior rims.

Measurements of expansion for each region pair were done in the same
way as for the entire remnant. We assumed a common expansion center of
coordinates for all pairs for Epochs I and II, but the measured
expansion rates are insensitive to this assumption. Slightly different
region positions and areas (but not shapes) were used for Epochs I and
II, to approximately match the measured expansion rates.  
The common expansion centers of (J2000)
coordinates are R.A.~$17^h48^m45.^s639 \pm 0.^s004$ 
($17^h48^m45.^s629 \pm 0.^s002$),
Decl.~$-27^\circ10'06."85 \pm 0."06$ ($-27^\circ10'06."96 \pm 0."03$) for 
Epoch I (II).  
A small ($0.''17$) but
significant difference between these centers suggests that the
coordinates of the corresponding reference observations ID6708 and
ID10112 are slightly misaligned. The magnitude of this shift is
consistent with the {\sl Chandra} external astrometric errors
\citep[mean error of $0.''16$;][]{rots09}.

\begin{deluxetable*}{lccccccccc}
\tablecolumns{10}
\tablewidth{0pc}
\tablecaption{Expansion Rates and Fluxes}

\tablehead{
\colhead{Region}  & Expansion Rate & $\dot{S}_{\rm 2007}$\tablenotemark{a} & $\dot{S}_{\rm 2009}$\tablenotemark{a} & $N_H$\tablenotemark{b} & $\Gamma$\tablenotemark{c} &$F_{\rm 2007}$\tablenotemark{d}  & $F_{\rm 2009}$\tablenotemark{d}  &$F_{\rm 2011}$\tablenotemark{d} & $\dot{F}$\tablenotemark{e}  \\
\colhead{} &(\%\ yr$^{-1}$) & \multicolumn{2}{c}{(\%\ yr$^{-1}$)} & ($10^{22}$ cm$^{-2}$) & &\multicolumn{3}{c}{($10^{-13}$ ergs cm$^{-2}$ s$^{-1}$)} & (\%\ yr$^{-1}$)}
\startdata

Total & 0.589 & -0.2 & 0.4 & 7.25 & 2.40 & 27.34  & 28.79  & 29.73 & 1.9 \\
& (0.573, 0.605) & (-0.7, 0.2) & (-0.1, 0.9)
& (7.16, 7.34) & (2.37, 2.43)
&(26.75, 27.93) &(28.50, 29.08) &(29.57, 29.99) & (1.5, 2.3)  \\

Outer SE-NW pair & 0.523 & 0.0 & 1.9 & 7.90 & 2.13 & 2.72 & 2.87 & 2.89 & 0.9 \\
& (0.487, 0.560) & (-1.5, 1.7) & (0.2, 3.6)
& (7.60, 8.21) & (2.04, 2.23)
&(2.61, 2.83) &(2.81, 2.92) &(2.86, 2.91) & (-0.3, 2.2) \\

Middle SE-NW pair & 0.616 & -0.4 & 1.0 & 7.58 & 2.34 & 9.16 & 9.75 & 10.05 & 1.9\\
& (0.592, 0.640) & (-1.1, 0.4) & (0.2, 1.8)
& (7.45, 7.72) & (2.29, 2.38)
&(8.97, 9.35) &(9.66, 9.84) &(10.00, 10.09) & (1.3, 2.5) \\

Inner SE-NW pair & 0.842 & 0.3 & -0.3 & 7.02 & 2.38 & 3.56 &3.66 &3.91 & 2.8 \\
& (0.783, 0.898) & (-0.9, 1.6) & (-1.6, 1.0)
& (6.82, 7.23) & (2.31, 2.45)
&(3.45, 3.70) &(3.60, 3.71) &(3.88, 3.94) & (1.8, 3.8) \\

N-S pair & 0.576 & -0.2 & -0.2 & 7.09 & 2.51 & 8.23 &8.65 &9.00 & 2.1 \\
& (0.544, 0.609) & (-1.0, 0.7) & (-1.1, 0.6) 
& (6.96, 7.23) & (2.46, 2.56)
&(8.05, 8.41) &(8.56, 8.74) &(8.95, 9.05) & (1.4, 2.8)
\enddata
\tablecomments{\ Expansion rates and fluxes in odd rows, 90\% confidence 
limits in even rows.}
\tablenotetext{a}{Surface brightness change.}
\tablenotetext{b}{Hydrogen column density.}
\tablenotetext{c}{Power-law photon index.}
\tablenotetext{d}{Absorbed flux in the 1--7 keV energy range.}
\tablenotetext{e}{Flux rate increase.}
\label{expansionfluxrates}
\end{deluxetable*}

\begin{figure}
\epsscale{1.2}
\plotone{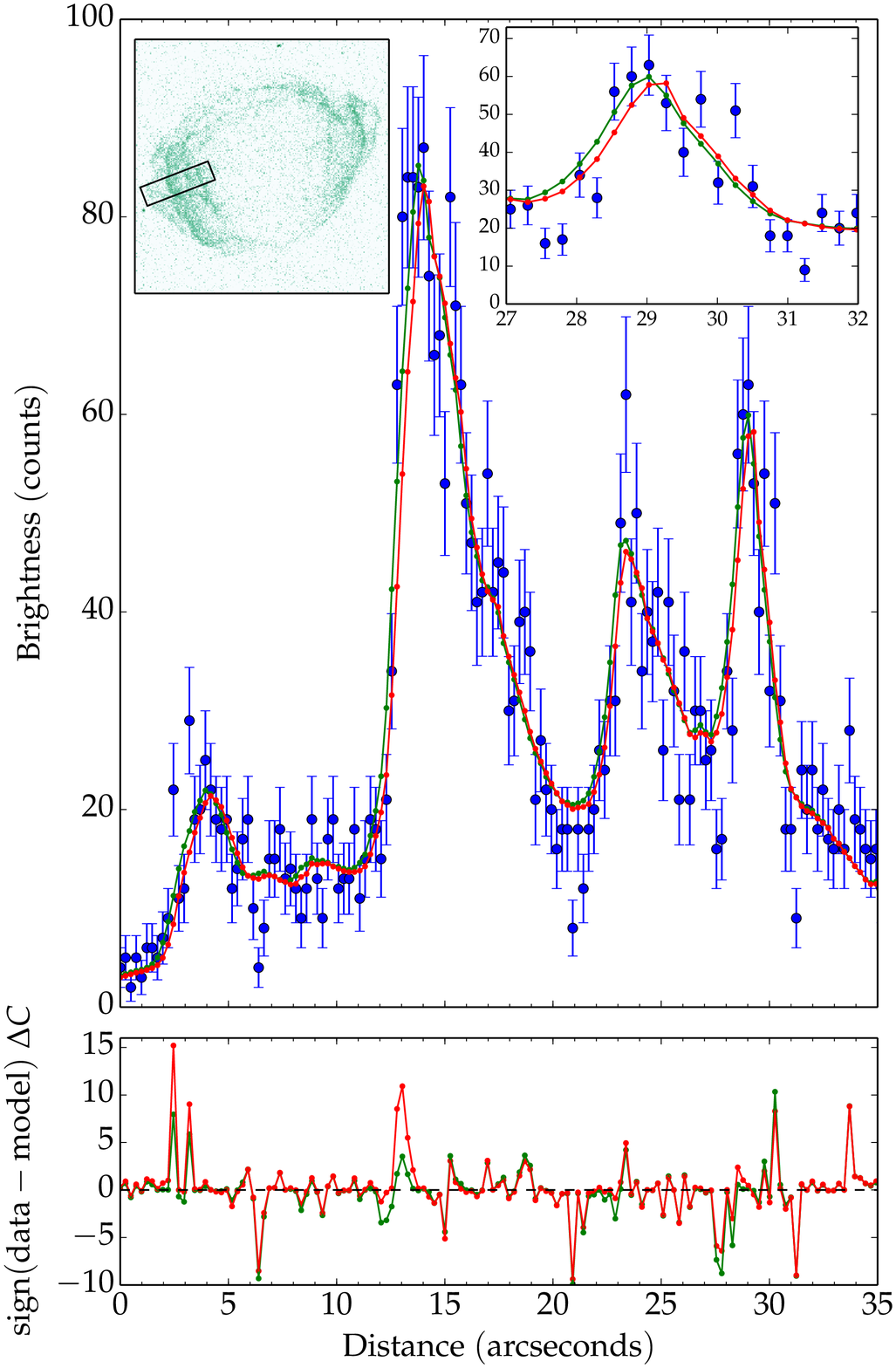}
\caption{Top: SE profile from 2009 (blue) along position shown on 2009 image 
(left inset), together with model profiles corresponding to best-fit 
expansion rates of $0.62$\%\ yr$^{-1}$ (green) and $0.84$\%\ yr$^{-1}$ (red) 
for the middle and inner rims (see Figure \ref{2xim} and 
Table \ref{expansionfluxrates}). Right inset: 
Zoomed view of the inner rim. Bottom: Contribution to C-statistic. 
Horizontal scale is distance along the profile, measured from E to W. 
\label{eastprofile}}
\end{figure}

\begin{figure}
\epsscale{1.2} 
\plotone{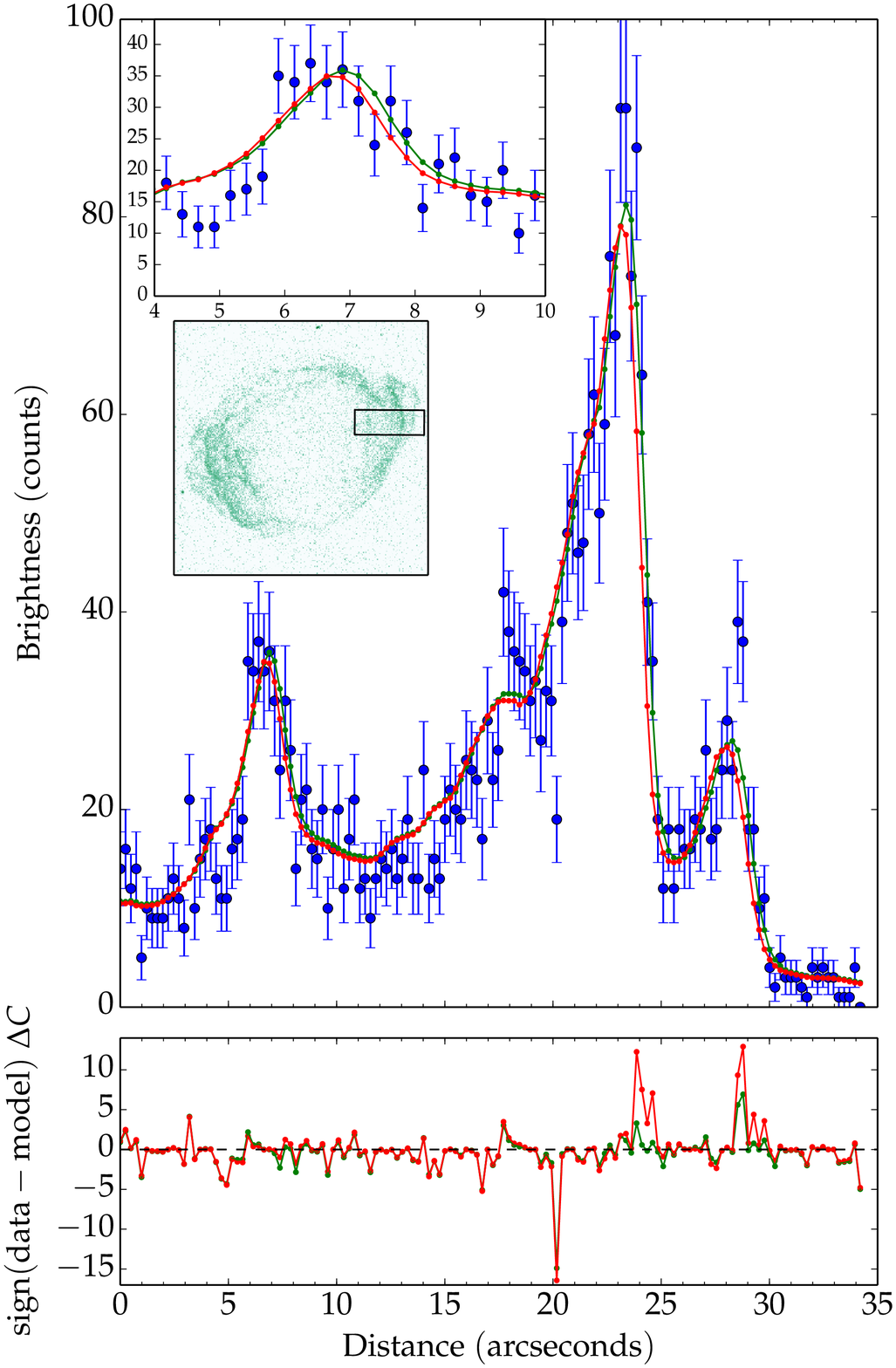}
\caption{NW profile from 2009. 
See caption to Figure \ref{eastprofile} for explanation.
\label{westprofile}}
\end{figure}

The measured expansion rates strongly deviate from uniform expansion
(Table \ref{expansionfluxrates}).  Expansion rates increase inward by
about 60\%\ along the bright SE-NW axis, ranging from $0.52\% \pm
0.03$\%\ yr$^{-1}$ for the outer ears to $0.84\% \pm
0.06$\%\ yr$^{-1}$ for the inner rims. The bright rims in the middle
expand slightly faster ($0.616\% \pm 0.024$\%\ yr$^{-1}$) than the
ears, but even this small difference is statistically highly
significant. The brightness-weighted linear displacement is 
$0.''29$ yr$^{-1}$ for 
all three rims. The N--S expansion is intermediate ($0.58\% \pm
0.04$\%\ yr$^{-1}$) between the bright middle rims and the ears, with  
the average displacement of $0.''23$ yr$^{-1}$. We
demonstrate this differential expansion for two
representative profiles in SE and NW based on the Epoch II
observations (Figures \ref{eastprofile} and \ref{westprofile}). 
As is seen most clearly in the close-up inserts, the 
expansion rate (green) that matches the bright middle rims 
is too slow (too little shrinkage) for the inner rims, while the 
faster expansion 
rate (red) that matches the inner rims is too fast for the 
middle and outer rims.

Contributions to C-statistic along the profiles quantify the fit
quality.  Systematic deviations are present even for the best-fit
models, perhaps due to spatial variation of the {\sl Chandra}
point-spread function, smaller-scale spatial variations in
expansion rate, or projection effects.  Strong counting
noise limits the accuracy of
the expansion measurements.  However, 
these profiles are merely illustrative; the results in Table 1 are
derived from the expansion of the entire regions shown in 
Figure \ref{2xim}.

\section{Flux Increase} \label{fluxincrease}

The spatially-integrated spectra of \object{G1.9+0.3} from each epoch
were jointly fit, together with their background spectra, 
with an absorbed power law in the 1 -- 9 keV energy
range. The power-law index and absorbing column density were
assumed constant in time, but the fluxes at each epoch were left
free. 
There is good agreement with previous measurements for Epoch I, 
but the newly-determined Epoch II flux is
2.5\%\ larger, in disagreement with the previous measurement also
reported in Paper V, presumably due to updates in the ACIS S3
calibration that were applied to all three datasets. 

Spatially-integrated fluxes are clearly increasing with time (Table
\ref{expansionfluxrates}). 
A likelihood ratio test 
reveals that the linear flux increase is consistent with the
individual flux measurements.
The measured flux increase is $1.9\% \pm 0.4$\%\ yr$^{-1}$, in
agreement with the previous, less accurate value of $1.7\% \pm 1.0\% $
yr$^{-1}$ (Paper V).

We also measured fluxes and rates of flux increase for the region
pairs shown in Figure \ref{2xim}, in the same way as for the
spatially-integrated fluxes except that the background contribution
was scaled down (rather than fit again) by the region/total area
ratios from the global fit.  
For each region
pair, the measured rate of flux increase is consistent with the
spatially-integrated rate.  There is an apparent trend in brightening
rate with radius, but uncertainties are large, due to uncertainties in
expanding the regions and to dust scattering of emission from brighter
regions (Paper III), as well as to large measurement errors.  We
conclude that evidence for spatial variations in the rate of 
flux increase is weak, although it should be the subject of
future investigations.

\section{Discussion}

The nonuniform expansion we observe for our three pairs of regions can
be rephrased as large differences in expansion age ($t_{\rm exp}
\equiv \Delta t\, R/\Delta R$), in the sense that $t_{\rm exp}$ is
largest (greatest deceleration) for the 
slower-expanding outermost material (see Figure
\ref{2xim} and Table \ref{expansionfluxrates}). The outermost ears,
bright rim emission in the middle, and distinct interior rims all have 
measurably
different expansion ages: 190 yr, 160 yr, and 120 yr.  We define
expansion indices $m \equiv d\, \ln r/d\, \ln t$ so that a feature at
radius $r$ obeys $r \propto t^m$ 
(note that $d\, \ln r/d\, \ln t = d\, \ln R/d\, \ln t$, where $R$ is the 
projection of the true radius $r$ onto the plane of the sky).
Then the true remnant age 
$t = m_{\rm fw} t_{\rm exp}$, where $m_{\rm fw}$ is the forward-shock expansion
index. 
Since we only have an estimate of 100 yr for the true 
remnant age (Paper V), we can only determine relative 
$m$ values.

The large spread in expansion ages between the inner rims and the ears
implies a large deceleration of the forward shock, $m_{\rm fw}
\lesssim 0.6$, significantly stronger than expected ($m_{\rm fw} \sim
0.7$) in the models with smoothly varying density distributions for
ejecta and ambient medium considered in Paper V. 
But the more
fundamental problem is that, contrary to observations, $m$ in these 
models typically varies only
slightly (if at all) from the reverse shock to the blast wave, and it
increases instead of decreasing with radius
\citep[][hereafter DC98]{chevalier82,dwarkadas98}. 
Either the ejecta or the ambient
medium density distribution (or both) must be very different from the
slowly-varying density distributions considered so far. Various
possibilities include substantial clumping or sudden density jumps
within the ejecta or the ambient medium.

Fine-scale clumping within the ejecta is unlikely to explain faster than 
expected expansion of the inner rims, however. Much like the outer rims, 
the inner rims consist of cohesive and
continuous filaments in contrast to the much more clumpy north rim
(Figure \ref{2xim}) where the ejecta emit most strongly in thermal
X-rays (Paper VI). Another explanation involves a rapid deceleration of the
blast wave upon a recent encounter with a moderate density jump in the
ambient medium. In order to account for about 60\%\ -- 70\%\ slower
expansion of the middle and outer rims relative to the inner rims
(Table \ref{expansionfluxrates}), a factor of 3 -- 6 density jump is
required \citep[from eqs.~4 and 6 in][]{borkowski97}.  A density jump
of this magnitude is consistent with a wind termination shock.  If
this were true, the SN progenitor must have been losing mass in a
strong and asymmetric stellar wind.  We estimate the wind parameter
($\equiv \dot{M}/v_w$, where $\dot{M}$ is the mass-loss rate and $v_w$
is the wind speed) at $4 \times 10^{-6} M_\odot\ {\rm yr}^{-1}/1000$
km s$^{-1}$, using the analytic wind thin-shell solution for an
exponential ejecta density distribution from Paper V with the remnant's 
radius of 2 pc and undecelerated age of 120 yr, and assuming
a standard thermonuclear explosion with kinetic energy $10^{51}$ ergs and
ejected mass equal to the Chandrasekhar mass. Such a strong wind favors 
a single-degenerate progenitor \citep{hachisu96}. The deceleration is 
$m=0.88$ in this model, suggesting that the explosion occurred sometime in the 
first decade of the 20th century. In this scenario, it is
difficult to understand the origin of the strong north-south asymmetry seen at
radio wavelengths (Figure \ref{2xim}) and in the spatial distribution
of thermal X-ray emission. As discussed in Paper VI, a strongly
asymmetric Type Ia explosion provides the best explanation for this
asymmetry.

\begin{figure}
\epsscale{1.2} 
\plotone{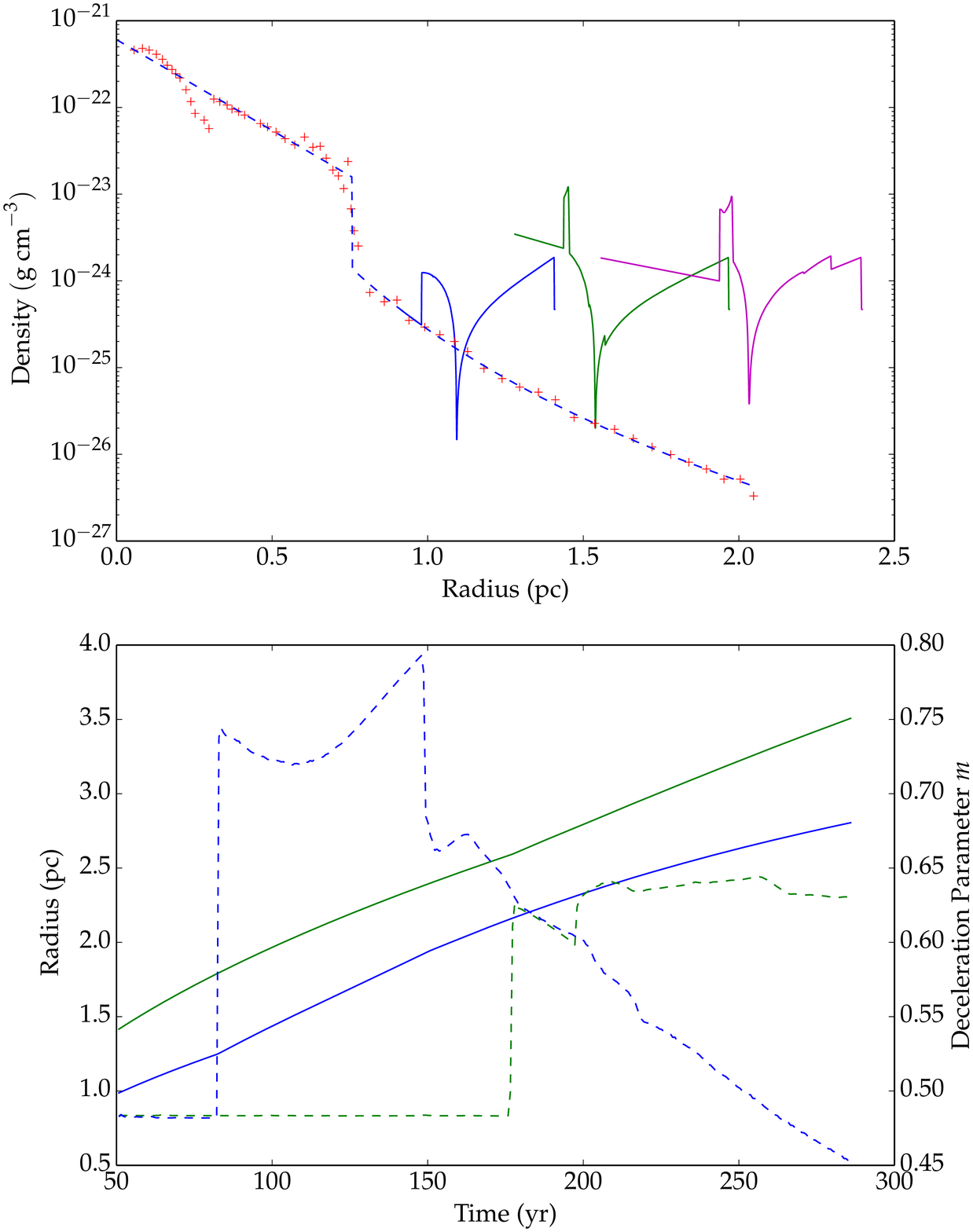}
\caption{Top: Density vs.~radius in 1-D hydrodynamical simulations with 
a composite power-law-exponential ejecta model and the uniform ambient medium 
with $n_0=0.2$ cm$^{-3}$ at 50 yr (solid blue line), 100 yr 
(green), and 150 yr (magenta). Undecelerated (freely expanding) ejecta at 
50 yr are also 
shown: composite model (dashed line), and the PDDEL1 model \citep{dessart14} 
($+$ signs). Bottom: Radii of the reverse and forward shocks (in blue and 
green solid lines), and their deceleration parameters $m$ (dashed lines). 
The reverse shock expansion is faster than the blast wave expansion during 
a long ($\sim 100$ yr) period of time. 
\label{hydro}}
\end{figure}

Unlike for smooth ejecta density profiles (such as resulting 
from delayed-detonation explosions),
it is possible to obtain less deceleration inward ($m_{\rm fw}
< m_{\rm rev}$) in 1-D numerical models for SNe Ia expanding into 
a uniform ambient
medium for ejecta profiles with substantial structure 
\citep[e.g., the PDDe model in ][]{badenes03}. Prominent density structures 
in the outer ejecta layers are present in the deflagration models 
(e.g., the W7 model of \citealt{nomoto84}), sub-Chandrasekhar 
explosions \citep[DC98;][]{badenes03}, and in
pulsating 
delayed-detonation (PDD) models \citep[e.g.,][see their PDDEL1 model 
plotted in Figure \ref{hydro}]{dessart14}.  
This density profile 
can be satisfactorily approximated by a
power law $\rho \propto r^{-n}$ with $n=5.8$ above the transition
velocity $v_{\rm tr}=14,800$ km s$^{-1}$, and by the exponential model
of DC98
at lower velocities. At $v_{\rm tr}$, there is
a large ($11.4$) density jump. 

In order to investigate how a density jump affects speeds of the reverse and 
forward shocks,
we performed 1-D hydrodynamical simulations with the VH-1 hydrocode
\citep[for a recent description of this code, see][]{warren13} for
this composite power-law-exponential ejecta model, 
We discuss here results for a preshock density $n_0$ of
$0.2$ cm$^{-3}$, matching the preshock density found in a young Type Ia 
\object{SNR 0509-67.5} in the Large Magellanic Cloud \citep{williams11}, 
but they can be scaled to any value of $n_0$ (DC98).
Prior to the impact of the reverse shock with
the density jump, the density profile can be described by the
self-similar solutions of \citet{chevalier82} with 
$m_{\rm fw}=m_{\rm rev}=(n-3)/n=0.48$ (see the density profile at 50 yr in Figure
\ref{hydro}). The reverse shock arrives at the density jump at $t = 83
(n_0/0.2\ {\rm cm}^{-3})^{-1/3}$ yr with velocity $v_s = 3v_{\rm
  tr}/n=0.52v_{\rm tr}$ (in the frame of reference moving with the
ejecta), and then splits into transmitted and reflected shocks. The
transmitted shock is the decelerated reverse shock
(inward-facing). Its velocity is $v_t=(\beta/\delta)^{1/2}v_s$ where
$\beta$ is the pressure enhancement that depends only on 
the density jump $\delta$ and
varies between 1 and 6 \citep[see equation 6 in][]{borkowski97}. In
the rest frame of the explosion, deceleration of the transmitted
reverse shock becomes $m_{\rm rev}=1-3(\beta/\delta)^{1/2}/n$. With
$\delta=11.4$, the overpressure is $\beta=2.76$, and $m_{\rm rev}$
increases to $0.75$, in good agreement with hydrodynamical simulations
(Figure \ref{hydro}). The reflected shock propagates first back into
the shocked ejecta, and then into the shocked ambient medium. At 100 yr, it
has already passed through the low-density contact discontinuity that
separates the shocked ejecta from the shocked ambient medium, and can
be seen in Figure \ref{hydro} as a small density discontinuity near
the contact discontinuity. The reflected shock strengthens with time
(see density profile at 150 yr), and eventually merges with the blast
wave at 175 yr, resulting in an abrupt increase of $m_{\rm fw}$ from
0.48 to 0.62. By this time, $m_{\rm rev}$ has already decreased to
about the same value after a transient phase following the sudden
deceleration of the reverse shock. This decrease continues in the
subsequent evolution, while $m_{\rm fw}$ stays about constant, so that
$m_{\rm fw} > m_{\rm rev}$.

During an extended period of time comparable with a young Type Ia
remnant's age, PDD explosions show a more rapid expansion of the
reverse shock than the forward shock (Figure \ref{hydro}). While this
still might be a reasonable interpretation for \object{G1.9+0.3},
published PDD models fail to match its properties in detail. Assuming the
ejecta structure of the composite model just described and identifying
the blast wave with the bright middle rims at $r = 2.04 (d/8.5\ {\rm kpc})$
pc expanding at 0.62\%\ yr$^{-1}$, the inferred preshock density
$n_0$ and free expansion ejecta velocity at the reverse shock are
$0.08 (d/8.5\ {\rm kpc})^{-5.8}$ cm$^{-3}$ and 
$18,000 (d/8.5\ {\rm kpc})$ km s$^{-1}$. At 8.5 kpc, $n_0$ is several 
times higher than estimated in Paper V, 
but the ejecta velocity is consistent with previous
estimates. \object{G1.9+0.3} is dynamically too young, however, for 
the reverse
shock to have reached the density jump at $v_{tr} = 14,800$ km
s$^{-1}$. No published PDD model has $v_{tr}$ as high as $18,000$ km
s$^{-1}$, with the highest ($16,000$ km s$^{-1}$) in the PDDa
model \citep{badenes03}. Among models considered by \citet{badenes03},
only the sub-Chandrasekhar model SCH has higher $v_{tr}$ at $18,500$
km s$^{-1}$.  Furthermore, sub-Chandrasekhar explosions produce large (up to
several $\times 10^{-3} M_\odot$) amounts of radioactive $^{44}$Ti
\citep[e.g.,][]{woosley11}, while the strength of the (possible)
$^{44}$Sc line implies that at most $\sim 10^{-5} M_\odot$ of $^{44}$Ti
was expelled by the SN that produced \object{G1.9+0.3}
\citep{borkowski13a}. 

Only the outermost ($v \gtrsim 18,000 (d/8.5\ {\rm kpc})$ km s$^{-1}$)
ejecta have been shocked so far in \object{G1.9+0.3} 
\citep[velocities might be as high as $v \gtrsim 21,000$ km s$^{-1}$ if 
$d \ge 10$ kpc;][]{roy14}. Discrete absorption features with such extreme
velocities are commonly seen in Type Ia SN spectra
\citep[e.g.,][]{childress14}, perhaps indicating the presence of dense
shells of material \citep{tanaka06}. Their origin 
is unknown, but possible explanations include a collision of
ejecta with circumstellar medium or clumping within the
ejecta. Fast dense shells with an order-of-magnitude density contrast
relative to overlying tenuous ejecta are expected to decelerate the
reverse shock just like the density jump in PDD models. This
might be the explanation for the fast expansion of the reverse shock
in \object{G1.9+0.3}. 

The flux increase may be due to global effects such as an increase in
magnetic-field strength, which would raise radio fluxes as well,
and/or an increase in the maximum energy to which electrons are
accelerated, which could raise X-ray fluxes even with constant radio.
The radio flux of \object{G1.9+0.3} is rising at 1 -- 2\% yr$^{-1}$
\citep{green08,murphy08}; our ongoing JVLA observations should reduce the
uncertainty to determine if this rate is consistent with the X-ray
rate.  A higher X-ray than radio rate of increase would require an
increasing maximum energy, a very interesting result.  Spatial
variations contain additional information.  One might naively expect
the greatest brightness increase at the location of greatest
deceleration (thermalization of kinetic energy); our data suggest but
do not compel this conclusion.  

\object{G1.9+0.3} offers us a unique
opportunity to study the dynamics and spatial distribution of the
outermost ejecta of a likely Type Ia SN, and the process of particle 
acceleration in very fast collisionless shocks. 
Our current understanding of
its dynamics and of spatial variations in the flux rate increase is limited by 
the short time baseline of the Epoch I --
III observations and the poor signal to noise of the 2007 and 2009
{\sl Chandra} datasets. 
Further {\sl Chandra} observations will 
allow for significant advances in our understanding of 
\object{G1.9+0.3}.

This work was supported by NASA through {\sl Chandra} General Observer
Program grants SAO G01-12098A and B.

\end{document}